\documentstyle[aps,prl,epsf]{revtex}
\draft
\newcommand{\singlefig}[2]{
\begin{center}
\begin{minipage}{#1}
\epsfxsize=#1
\epsffile{#2}
\end{minipage}
\end{center}}
\newenvironment{figcaption}[2]{
\vspace{0.3cm}
\refstepcounter{figure}
\label{#1}
\begin{center}
\begin{minipage}{#2}
\begingroup \small FIG. \thefigure: }{
\endgroup
\end{minipage}
\end{center}}
\begin{document}

\wideabs{
\title{Possible direct method to determine THE radius of a star \break
from the spectrum of gravitational wave signals}
\author{
Motoyuki Saijo$^{1}$
and
Takashi Nakamura$^{2}$
}
\address{
$^{1}$
Department of Physics, Waseda University,
3-4-1 Okubo, Shinjuku, Tokyo 169-8555, Japan
}
\address{
$^{2}$
Yukawa Institute for Theoretical Physics, Kyoto University,
Kyoto 606-8502, Japan
}
\date{Received 14 February 2000}
\maketitle

\begin{abstract}
We computed the spectrum of gravitational waves from a dust disk
star of  radius $R$ inspiraling into a Kerr black hole of mass $M$ and
specific angular momentum $a$.  We found that when $R$ is much
larger than the wave length of the quasinormal mode, the spectrum has
several peaks and the separation of peaks $\Delta\omega$ is
proportional to $R^{-1}$ irrespective of $M$ and $a$. This suggests
that  the radius of the star in coalescing binary black hole - star
systems may be determined directly from the observed spectrum of
gravitational wave. This also suggests that   the spectrum of the
radiation  may give us important information in gravitational wave
astronomy as in optical astronomy.
\end{abstract}
\pacs{PACS numbers: 04.30.Db, 04.25.Nx}
}

A number of worldwide projects on the direct detection of
gravitational waves using  laser interferometers such as LIGO, VIRGO,
GEO600, TAMA300, and  LISA as well as several projects using
resonant type detectors are in progress \cite{Thorne}. One of the most
important sources of gravitational waves for these detectors is
coalescing binary black hole (BH) - star systems. In the inspiral
phase, the radius of the star has  little effect on gravitational waves
so that the star can be treated as a point particle. Comparing the
waveform of the inspiral phase with a theoretical template using
matched filtering techniques, we may determine each mass and spin of
the star and BH, respectively \cite{Thorne}.  While, in the final merging
phase, the effect of the radius is important  and some information on
the radius of the star may be extracted from gravitational wave
signals. However, this merger phase is much less well understood than
the inspiral one.

In this Letter, as a model of this final merging phase, we computed the
spectrum of gravitational waves from a dust disk star of  radius $R$
inspiraling into a Kerr BH of mass $M$ and specific angular momentum
$a$. We found that when $R$ is much larger than the wave length of the
quasinormal mode (QNM) of Kerr BH, the spectrum has several peaks
and the separation of peaks $\Delta\omega$ is proportional to $R^{-1}$
irrespective of $M$ and $a$.  This suggests that the radius of the star
in coalescing binary BH - star systems may be determined directly
from the observed spectrum of gravitational waves without any
assumption to equation of state.  \footnote{Vallisneri \cite{Vallisneri}
recently argued that the radius of the star may be determined if the
frequency of gravitational waves at the tidal disruption is obtained
observationally.  However, his method needs some information about
the equation of state \cite{Shibata} to know  the frequency at the tidal
disruption.}

First, let us consider gravitational waves from a particle plunging into
a Kerr BH of mass $M$ and specific angular momentum $a$. The Fourier
component of the radial wave function of gravitational waves
$X_{lm\omega}(r^*)$ obeys the Sasaki-Nakamura equation (Eq. (2.28) of
\cite{SN2}) given by
\begin{equation}
\left[ \frac{d^{2}}{dr^{*2}} - F(r^*)\frac{d}{dr^*} - U(r^*)  \right]
X_{lm\omega}(r^*) = S_{lm\omega}(r^*),
\label{eqn:GeneRW}
\end{equation}
where $r^{*}$, $S_{lm\omega}(r^{*})$, $F(r^{*})$ and $U(r^{*})$ are the
tortoise coordinate of Kerr BH, the source term from the test particle
of mass $\mu$ (Eq. (2.29) of \cite{SN2}) and two potential functions
(Eqs. (2.12a) and (2.12b) of \cite{SN2}), respectively.  To solve
$X_{lm\omega}(r^{*})$  using a Green function method, we need two
independent homogeneous solutions  whose boundary conditions are
given by
\begin{eqnarray*}
X_{lm\omega}^{\rm in(0)}(r^*) & = & \left\{
\begin{array}{ll}
e^{-i k r^{*}} &  r^{*}\rightarrow -\infty \\
A_{lm\omega}^{\rm in}e^{-i \omega r^{*}} + A_{lm\omega}^{\rm out}e^{i
\omega r^{*}} &  r^{*}\rightarrow \infty
\end{array}
\right. ,\\
X_{lm\omega}^{\rm out(0)}(r^*) & = & \left\{
\begin{array}{ll}
B_{lm\omega}^{\rm in}e^{-i k r^{*}} + B_{lm\omega}^{\rm out}e^{i k
r^{*}}
& r^{*}\rightarrow - \infty \\
e^{i \omega r^{*}} & r^{*}\rightarrow \infty
\end{array}
\right. ,
\end{eqnarray*}
where $k=\omega-ma/[2 (M+\sqrt{M^{2}-a^{2}})]$.  Then, the
inhomogeneous solution to Eq. (\ref{eqn:GeneRW}) becomes
\begin{eqnarray*}
X_{lm\omega}(r^{*}) & = &
X_{lm\omega}^{\rm in(0)}
\int_{r^{*}}^{\infty} \frac{S_{lm\omega}(r^{*})}{W}
X_{lm\omega}^{\rm out(0)}~dr^{*}
\nonumber \\
&&
+ X_{lm\omega}^{\rm out(0)}
\int_{-\infty}^{r^{*}} {S_{lm\omega}(r^*)\over {W}}
X_{lm\omega}^{\rm in(0)}~dr^{*},
\end{eqnarray*}
where $W$ is the Wronskian,
\begin{equation}
W  \equiv X_{lm\omega}^{\rm in(0)}
\frac{dX_{lm\omega}^{\rm out(0)}}{dr^{*}} -
X_{lm\omega}^{\rm out(0)}
\frac{dX_{lm\omega}^{\rm in(0)}}{dr^{*}}
.
\end{equation}
The asymptotic behavior  of the radial wave function
$X_{lm\omega}(r^{*})$ is given by
\begin{eqnarray}
X_{lm\omega}(r^{*}) & = &
A_{lm\omega} e^{i\omega r^{*}},
\label{eqn:RWsol}\\
A_{lm\omega} & = & \int_{-\infty}^{\infty} {S_{lm\omega}(r^*)\over {W}}
X_{lm\omega}^{\rm in(0)}~dr^{*}
. \nonumber
\end{eqnarray}
Then, the energy spectrum and waveform of gravitational waves (Eq.
(3.6) and (3.10) of \cite{SSM}) are given by
\begin{eqnarray}
\left( \frac{dE}{d\omega} \right)_{lm\omega} & = &
8 \omega^{2}
\left| \frac{A_{lm\omega}}{c_{0}} \right|^{2}
\hspace{1cm}
(- \infty < \omega < \infty)
\label{eqn:dEdw}
,\\
h_{+}-ih_{\times} &=&
\frac{8}{r} \int_{-\infty}^{\infty} d\omega
e^{i \omega (r^{*} - t)}
\nonumber \\
&& \times
\sum_{l,m} 
\left[ 
  \left( \frac{A_{lm\omega}}{c_{0}} \right)
  \mbox{}_{-2}S_{lm}^{a\omega}(\theta)
  \frac{e^{i m \varphi}}{\sqrt{2\pi}}
\right]
,
\label{eqn:Waveform}
\end{eqnarray}
where $r$ is the coordinate radius from the center of BH,
$\mbox{}_{-2}S_{lm}^{a\omega}$ is the spin - 2 weighted spherical
harmonics, $c_{0}$ is a constant  given in \cite{SSM}.

Next, we consider gravitational waves from a dust disk
\cite{SSM,NS,HSW}. We note that a geodesic in the equatorial plane in
Kerr BH is characterized by  the specific energy ($\tilde{E}$) and
angular momentum ($\tilde{L}_{z}$). Let  $t=T(r)$ and $\phi=\Phi(r)$
express an orbit of the geodesic for given $\tilde{E}$  and
$\tilde{L}_{z}$. Then another geodesic for the same $\tilde{E}$  and
$\tilde{L}_{z}$ is expressed by  $t=T(r)+c_t$ and
$\phi=\Phi(r)+c_{\phi}$  where $c_{t}$ and $c_{\phi}$ are constants.
Specifically  the geodesic, whose location at $t=T(r_{0})$ is $r=r_{i}$
and $\phi=\phi_{i}$, is expressed by $t=T(r)+T(r_{0})-T(r_{i})$ and
$\phi=\Phi(r)+\phi_{i}-\Phi(r_{i})$.  It is possible to set a number of
particles to form  a disk of radius $R$ whose center is $r=r_{0}$ and
$\phi =\Phi (r_{0})$ at $t=T(r_{0})$.


Each particle in the disk emits the same gravitational waves as the
particle in the center of the disk except that the phase in $t$ and
$\phi$ is  different from the central particle  \cite{NS}. Then the
amplitude of the gravitational waves from the dust disk star is
computed as
\begin{eqnarray}
A_{lm\omega}^{\rm (disk)}
&=& f_{m\omega} A_{lm\omega}^{\rm (particle)}
,\\
f_{m\omega}
&=&
2 \frac{\mu}{S}
\int_{r_{0} - R}^{r_{0} + R} dr r \frac{\sin (m \phi_{0} (r))}{m}
\nonumber \\
&& \times
e^{i [ \omega [ T(r) - T(r_{0}) ] -
m [ \Phi(r) - \Phi(r_{0}) ] ]},
\label{eqn:DiskRadialFunction}
\\
\phi_{0}(r) &=&
cos^{-1} \frac{r^{2}+r_{0}^{2}- R^{2}}{2rr_{0}},
\end{eqnarray}
where $f_{m\omega}$ and $A_{lm\omega}^{\rm (particle)}$ are a form
factor and the amplitude of gravitational waves at infinity for a single

particle case, respectively.  $S$ is the normalization factor given by
\begin{eqnarray}
S &=&
2 \int_{r_{0} - R}^{r_{0} + R} dr r \phi_{0} (r).
\label{eqn:sphere}
\end{eqnarray}

In this Letter, we only show the result for $r_{0} = 10M$, $a/M = 0.9$,
$\tilde{E}=1$ and $\tilde{L}_{z}=2 M$.  In this case, the orbit rotates
about $\pi$ angle around the BH within the  radius of $r=r_{0}$, and
the velocity at $r=3M$, the location where the  particle might radiate
the characteristic QNM frequency, is $v \sim 0.48 c$, where $c$ is the
speed of light.  As for the choice of the parameter $r_{0}=10M$,  we
can refer the result of the relativistic Roche limit (Eq. (5.1) of
\cite{Shibata}) as
\begin{figure}
\begin{center}
\singlefig{7.3cm}{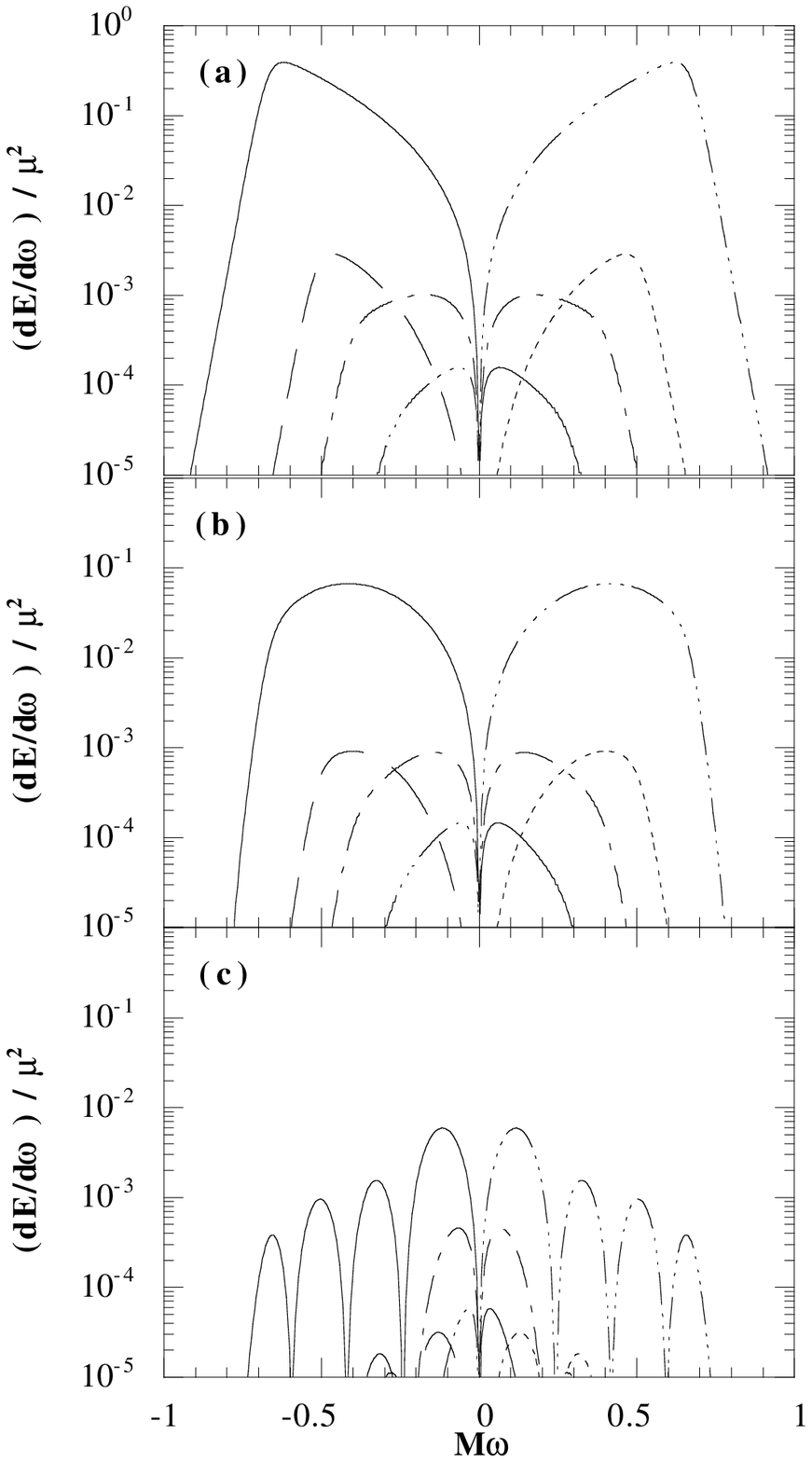}
\begin{figcaption}{fig:sp}{8.5cm}
Energy spectrum of gravitational waves from a dust disk star moving
on an equatorial plane in Kerr spacetime whose radius is set up at
$r_{0}=10M$ for the case of $a/M=0.9$, $\tilde{L}_{z} / M = 2$ ((a)
$R/M=0$ (test particle), (b) $R/M=1.56$, (c) $R/M=5.88$).  We only show
$l=2$ mode. Solid, dashed, dash-dotted, dotted, and dash-three dotted
lines denote the case of $m=-2$, $-1$, $0$, $1$, $2$, respectively.
\end{figcaption}
\end{center}
\end{figure}

\begin{equation} 
M < 4.64 M_\odot 
\left( \frac{1.4 M_\odot}{M_s} \right)^{1/2} 
\left( \frac{R}{10{\rm km}}\right)^{3/2}
\left( \frac{6M}{r_c} \right)^{3/2},
\end{equation}
where $M_s$ and $r_c$ are the mass of the star and the Roche radius,
respectively. For appropriate choice of parameters $r_{0}=10M$
corresponds to the Roche limit.

Fig. \ref{fig:sp} (a) and Fig. \ref{fig:gw} (a) show the characteristics of
gravitational waves from a test particle  plunging into a Kerr BH of
$a/M=0.9$. Fig. \ref{fig:sp} (a) shows the spectrum  for $l=2$ and Fig.
\ref{fig:gw} (a) shows the waveform $h_{+}$ observed from the
equatorial plane at infinity.  From Fig. \ref{fig:sp} (a), we see that
most of gravitational waves are radiated near the frequency of the
QNM since the QNM frequency for $a/M=0.9$ is ${\rm Re}~(M
\omega_{\rm QNM}) =0.7$ for $l=m=2$, while the imaginary part of QNM
(${\rm Im}~(M\omega_{\rm QNM})$) in this case is  $ 0.07$ (Fig. 1a of
\cite{Detweiler} and Fig. 3 (c) of \cite{Leaver}). Fig. \ref{fig:gw} (a)
shows clearly that there exists a typical ringing tail at  late time (Fig.
4 of \cite{KN1} and Fig. 4 of \cite{KN2}).  Fig. \ref{fig:sp} (b), (c) and
Fig. \ref{fig:gw} (b), (c) show the spectra and the waveform  from the
circular dust disk star inspiraling into a Kerr BH with $a/M=0.9$. The
radius of the disk is set up as $R/M=1.56$ for Fig. \ref{fig:sp} (b) and
Fig. \ref{fig:gw} (b) while  $R/M=5.88$ for Fig. \ref{fig:sp} (c) and Fig.
\ref{fig:gw} (c). For the case of Fig. \ref{fig:sp} (b), the peak of the
spectrum is located at $M\omega \sim 0.5$ which is quite different
from the QNM frequency ${\rm Re}~(M\omega_{\rm QNM}) \sim 0.7$
although the ringing tail can be seen in Fig. \ref{fig:gw} (b). When we
turn to look at Fig. \ref{fig:sp} (c), there are several peaks in the
spectrum and the ringing tail is very low or absent in the waveform of
Fig. \ref{fig:gw} (c).  This is a new finding. The reason for this
behavior is as follows. The energy spectrum
$(dE/d\omega)_{lm\omega}^{\rm (disk)}$ of gravitational waves from
the disk is expressed as
\begin{equation}
\left( \frac{dE}{d\omega} \right)_{lm\omega}^{\rm (disk)} \propto
\mid f_{m\omega}\mid ^2
\left( \frac{dE}{d\omega} \right)_{lm\omega}^{\rm (particle)},
\end{equation}
where $(dE/d\omega)_{lm\omega}^{\rm (particle)}$ is the spectrum
from the single test particle.  The spectrum from the test particle has
only one peak at the frequency $\omega_{\rm QNM}$ (Fig. \ref{fig:sp}
(a)) so that the square of the form factor $\mid f_{m\omega}\mid ^2$
is responsible for this behavior. To confirm this, we show $\mid
f_{m\omega}\mid ^{2}$ for $R/M=1.56$ (Fig. \ref{fig:form} (a)) and
$R/M=5.88$ (Fig. \ref{fig:form} (b)) and find that $\mid
f_{m\omega}\mid ^{2}$ is responsible for the behavior of the spectrum
in Fig. \ref{fig:sp} (b) and (c). The existence of several peaks in Fig.
\ref{fig:form} (a) and (b) can be understood by the approximate
estimation of $ f_{m\omega}$ as
\begin{equation}
 f_{m\omega}\propto \frac{\sin (\omega T'_{r=r_{0}} R)}{\omega},
\end{equation}
where $T'=dT(r)/dr$. The frequency where $f_{m\omega}$ takes zero is
\begin{equation}
\omega_{n} \sim \frac{(n+1)\pi}{\ T'_{r=r_{0}} R}~~n=0,1,2, \ldots ,
\label{eqn:EstimateF}
\end{equation}
which agrees quite well with the numerical results of Fig.
\ref{fig:form} (a) and (b). Equation (\ref{eqn:EstimateF}) suggests that
the separation of peaks of the spectrum $\Delta \omega$ may be in
proportion to $R^{-1}$. In Table \ref{tbl:EstimateR}, we show
$\Delta \omega$ for various $R/M$ and we found that
\begin{equation}
R=C\frac{1}{\Delta \omega}~ {\rm for}~ R/M \gg 1.6,
\label{eqn:EstimateR}
\end{equation}
where a constant $C$ is close to 1 in the present case.
In the physical unit, $\Delta \nu\equiv \Delta \omega/(2 \pi)$
is given by
\begin{equation}
\Delta \nu 
= 5{\rm kHz} \left( \frac{R}{10{\rm km}} \right)^{-1}
= 10{\rm Hz} \left( \frac{R}{5000{\rm km}} \right)^{-1},
\end{equation}
assuming that there are 4 peaks in the region of 
$\omega \in [0,\omega_{\rm QNM}]$ in the spectrum.  Therefore  for 
neutron stars and white dwarfs, the frequency band is within some
laser interferometers and resonant type detectors.
\begin{figure}
\begin{center}
\singlefig{7.3cm}{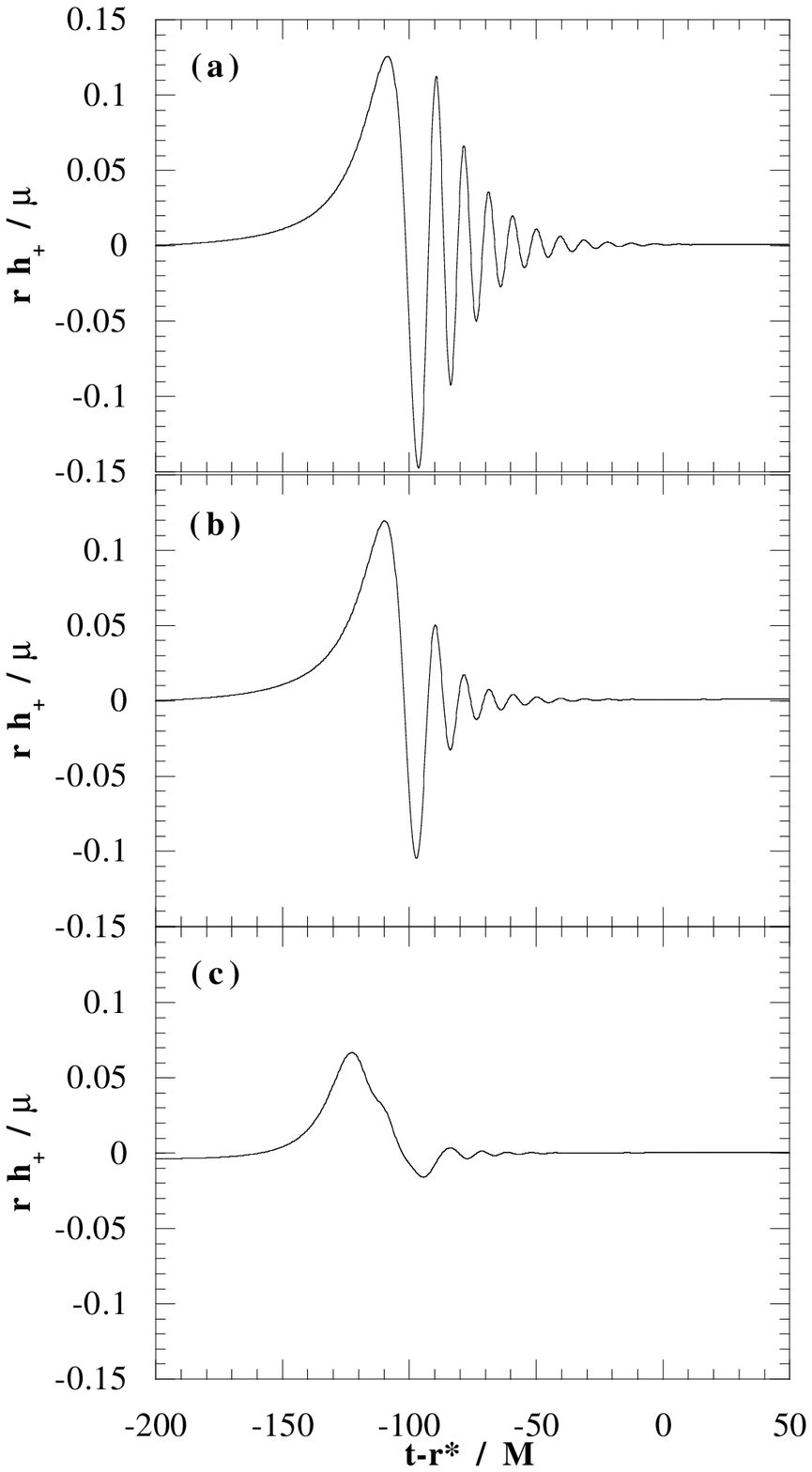}
\begin{figcaption}{fig:gw}{8.5cm}
Waveform of gravitational waves from a dust disk star moving on an
equatorial plane in Kerr spacetime whose radius is set up at
$r_{0}=10M$ for the case of $a/M=0.9$, $\tilde{L}_{z} / M = 2$ ((a)
$R/M=0$ (test particle), (b) $R/M=1.56$,  (c) $R/M=5.88$).  We set the
angle of the observer at infinity as $\theta = \pi / 2$, $\phi = 0$.  We
only show
$l=m=2$ mode to exclude $m$ mode coupling effect.
\end{figcaption}
\end{center}
\end{figure}

Equation (\ref{eqn:EstimateR}) strongly suggests that from the
observed value of $\Delta \omega$ in the spectrum of the gravitational
wave signals, we may determine the radius of the star if $ R/M \gg
1.6$.   From the physical point of view, the form factor $f_{m\omega}$
expresses the phase cancellation effect of gravitational waves
\cite{NS}, so that if $R < 1/\omega_{\rm QNM}$ gravitational waves
from the dust disk star is essentially regarded as that from the test
particle. \footnote{ We also confirmed that this phase cancellation can also be
seen in the 3D dust star assuming that each test particle moves the
constant polar angle with no orbital angular momentum.}  This suggests
that our proposal is  valid only if the inspiraling star is tidally
disrupted by the BH finally. It is instructive to note here that in
reality the form factor may bring us the information about the form of
the source, {\it radius}.

Here, we only showed the results for a special set of parameter such as
$r_{0}=10M$, $a/M=0.9$, $\tilde{E}=1$, $\tilde{L}_{z}=2M$.  To check
the robustness of our proposal, we also calculated the spectra from
the dust disk star inspiraling into  Kerr BH for the wide range of
parameters ($\tilde{L}_{z}$: spiraling case, $0 \leq a/M \leq 0.9$) and
found that there exist some peaks in the spectrum for $R \gg
1/\omega_{\rm QNM}$. In these cases we confirmed the relation
$R\sim 1/\Delta \omega $. In this Letter, we did not take into account
of the pressure and the self-gravity of the star so that it  is urgent to
confirm our proposal by full 3D numerical simulations including the
determination of the value of the constant $C$ in Eq.
(\ref{eqn:EstimateR}).  If the relation of Eq. (\ref{eqn:EstimateR}) is
confirmed irrespective of the equation of state, Eq.
(\ref{eqn:EstimateR}) can be adopted as  one of the direct methods to
determine the radius of the star from gravitational waves.  In any
case, it is quite possible that the spectrum of the gravitational waves
may give us important information in gravitational wave astronomy as
in optical astronomy.

\begin{figure}
\begin{center}
\singlefig{7.3cm}{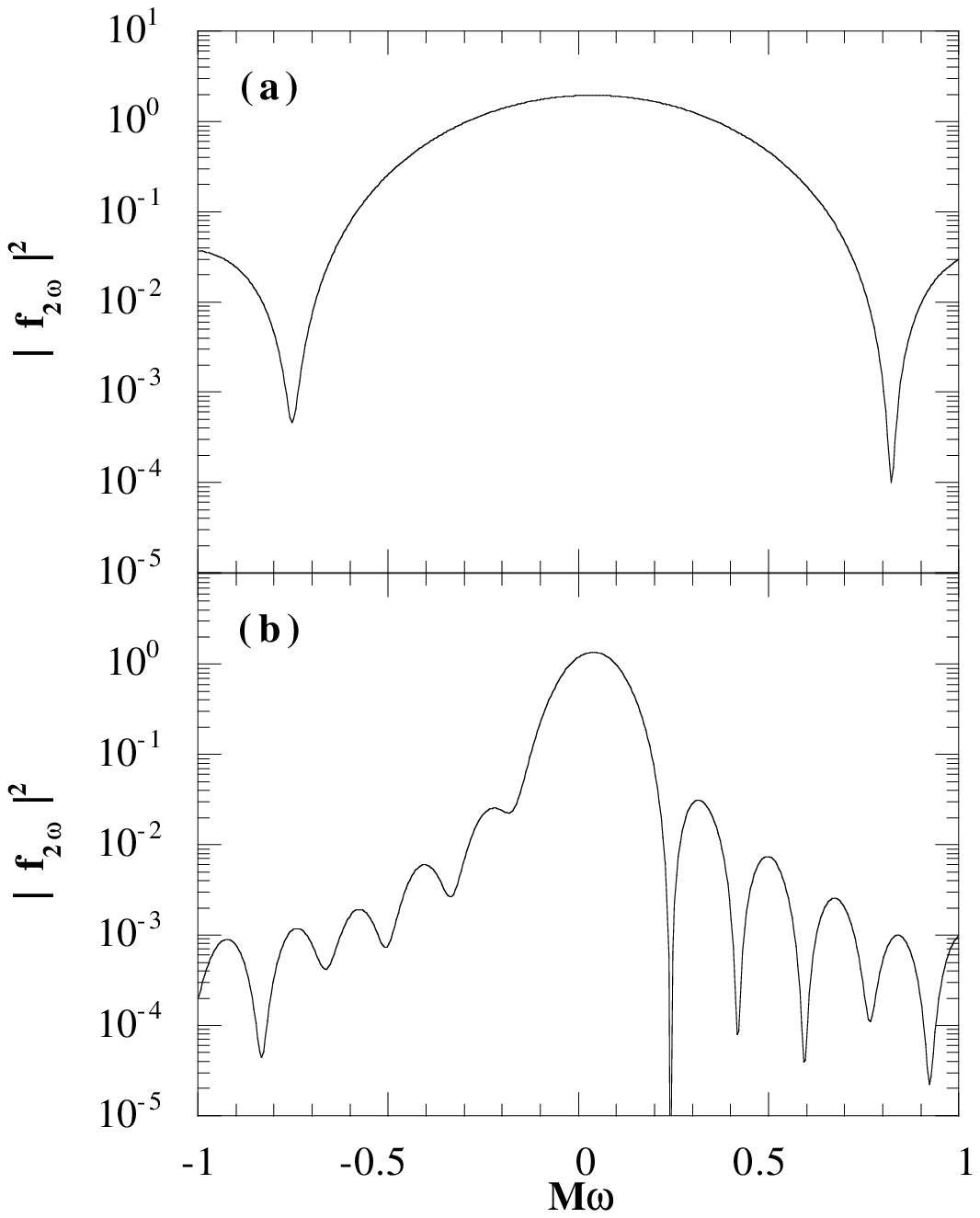}
\begin{figcaption}{fig:form}{8.5cm}
Form factor of the dust disk star moving on an equatorial plane in
Kerr spacetime whose radius is set up at $r_{0}=10M$ for the case of
$a/M=0.9$, $\tilde{L}_{z} / M = 2$ ((a) $R/M=1.56$,  (b) $R/M=5.88$).
We only show $l=m=2$ mode, because $m=2$ is a dominant mode for
above parameters in $l=2$. It is clear that the form factor is
responsible for the spectra in Fig. 1.
\end{figcaption}
\end{center}
\end{figure}

\begin{table}
\caption
{
Comparison with the characteristic length from the energy spectrum of
gravitational waves to the radius of the disk. $R$ denotes the
coordinate radius.
}
\begin{center}
\begin{tabular}{c c c}
$R/M$ &  $\Delta \omega$ &
$1/\Delta \omega$ \\
\hline
$3.09$ & $0.335$ & $2.99$
\\
$3.83$ & $0.270$ & $3.70$
\\
$4.54$ & $0.233$ & $4.29$
\\
$5.22$ & $0.200$ & $5.00$
\\
$5.88$ & $0.175$ & $5.71$
\\
\end{tabular}
\label{tbl:EstimateR}
\end{center}
\end{table}

M. S. would like to thank Kei-ichi Maeda for a continuous
encouragement.  He thanks Masaru Shibata for discussion.   He also
thanks the visitor system of Yukawa Institute for Theoretical Physics
and acknowledges Abhay Ashtekar, Lee Samuel Finn, Jorge Pullin,
Hisaaki Shinkai for their kind hospitality at The Pennsylvania State
University, where the part of this work was done. The Numerical
Computations are mainly performed by NEC-SX vector computer at
Yukawa Institute for Theoretical Physics, Kyoto University and
FUJITSU-VX vector computer at Media Network Center, Waseda
University.  This work was supported in part by a JSPS Grant-in-Aid 
(No. 5689) and by Grant-in-Aid of Scientific Research of the Ministry
of Education, Culture, and Sports, No.11640274 and 09NP0801.


%
\end{document}